\documentclass[
    ,final            
  ]
  {aipproc}

\layoutstyle{6x9}

\def\beq{\begin{equation}}
\def\eeq{\end{equation}}
\def\bea{\begin{eqnarray}}
\def\eea{\end{eqnarray}}
\def\nn{\nonumber}
\def\bge{\begin{equation}}
\def\ene{\end{equation}}
\def\bg{\begin{eqnarray}}
\def\en{\end{eqnarray}}

\def\q{{\bf q}}

\def\q{{\bf q}}
\def\r{{\bf r}}

\def\D0bar{\overline{D^0}}

\begin{document}

\title{$J/\Psi$ mass shift and $J/\Psi$-nuclear bound state}

\classification{21.85.+d,21.65Jk,21.65.Qr,24.85.+P}

\keywords{$J/\Psi$-nuclear bound state, in-medium $J/\Psi$ self-energy,
Quark-meson coupling model}

\author{K.~Tsushima}{
  address={CSSM, School of Chemistry and Physics,
University of Adelaide, Adelaide SA 5005, Australia}
}

\author{D.~H.~Lu}{
  address={Department of Physics, Zhejiang University, Hangzhou 310027, P.R.China}
}

\author{G. Krein}{
  address={Instituto de F\'{\i}sica Te\'orica, Universidade Estadual Paulista \\
Rua Dr. Bento Teobaldo Ferraz, 271 - Bloco II, S\~ao Paulo, SP, Brazil}
}

\author{A.~W.~Thomas}{
  address={CSSM, School of Chemistry and Physics,
University of Adelaide, Adelaide SA 5005, Australia}
}

\begin{abstract}
We calculate mass shift of the $J/\Psi$ meson in nuclear matter 
arising from the modification of $DD, DD^*$ and $D^*D^*$
meson loop contributions to the $J/\Psi$ self-energy.
The estimate includes the in-medium $D$ and $D^*$ meson masses consistently.
The $J/\Psi$ mass shift (scalar potential) calculated is negative (attractive),
and complementary to the attractive potential obtained 
from the QCD color van der Waals forces.
Some results for the $J/\Psi$-nuclear bound state energies are also presented.
\end{abstract}

\maketitle


\subsection{Introduction}

The properties of charmonia and charmed mesons in a nuclear medium (nuclei)
are still poorly known.
However, with the $12$~GeV upgrade of the CEBAF accelerator
at the Jefferson Lab and with the construction of the FAIR facility,
we expect tremendous progress in understanding
the properties of these mesons in a nuclear medium.
The new facilities will be able to produce low-momenta charmonia
and charmed mesons such as $J/\Psi$, $\psi$, $D$ and $D^*$
in an atomic nucleus.
One of the major challenges is to find appropriate kinematical conditions
to produce these mesons essentially at rest, or with small momentum
relative to the nucleus. Amongst many new possible experimental efforts,
to search for the $J/\Psi$-nuclear bound states may be one of
the most exciting ones.
Finding of such bound states would provide us with
evidence for the negative mass shift of the $J/\Psi$ meson discussed here.

There is a relatively long history for the studies of
$J/\Psi$ and $\eta_c$ binding in nuclei.
The original suggestion~\cite{Brodsky:1989jd} that multiple gluon
QCD van der Waals forces would be capable of binding a charmonium state,
estimated a binding energy as large as $400$~MeV in an $A=9$
nucleus. However, the same approach but taking into account
the nucleon density distributions in the nucleus,
Ref.~\cite{Wasson:1991fb} found a maximum of $30$~MeV
binding energy in a large nucleus.
Based on Ref.~\cite{Peskin:1979va},
which showed the mass shift of charmonium in nuclear matter
is possible to be expressed similar to the usual
second-order Stark effect due to the chromo-electric polarizability
of the nucleon, the authors of Ref.~\cite{Luke:1992tm}
obtained a $10$~MeV binding for $J/\Psi$ in nuclear matter,
in the limit of infinitely heavy charm quark mass.
Following the same procedure, but keeping the charm quark mass
finite and using realistic charmonium bound-state wave-functions,
Ref.~\cite{Ko:2000jx} found $8$~MeV binding energy for $J/\Psi$ in
nuclear matter.

There are some other studies on the $J/\Psi$ mass in nuclear medium.
QCD sum rules estimated a $J/\Psi$ mass decrease in nuclear matter
ranging from $4$ MeV to $7$ MeV~\cite{Klingl:1998sr,Hayashigaki:1998ey,Kim:2000kj},
while an estimate based on color polarizability~\cite{Sibirtsev:2005ex}
gave larger than $21$ MeV.
Since the $J/\Psi$ and nucleons have no quarks in common,
the quark interchange or the effective meson exchange
potential should be none or negligible to first order
in elastic scattering~\cite{Brodsky:1989jd}.
Furthermore, there is no Pauli blocking even at the quark level.
Thus, if the $J/\Psi$-nuclear bound states are formed, the signal for
these states will be sharp and show a clear narrow peak in energy dependence
of the cross section.

In this contribution we report on our study~\cite{psimass}
made for the mass shift of $J/\Psi$.
In addition we also report some results obtained for the
$J/\Psi$-nuclear bound states~\cite{psimesic}.

\subsection{$J/\Psi$ mass shift in symmetric nuclear matter}

\begin{figure}[htb]
\includegraphics[height=.15\textheight]{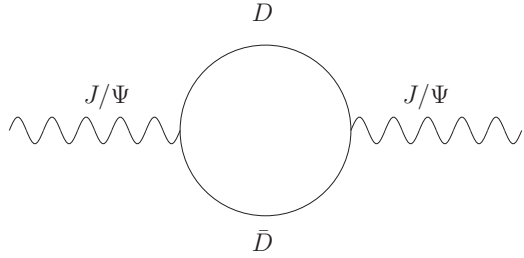}
\caption{$DD$-loop contribution to the $J/\Psi$ self-energy.}
\label{fig:loop}
\end{figure}

We study first the $J/\Psi$ meson mass shift in medium arising from 
the modification of $DD, DD^*$ and $D^*D^*$ meson loop
contributions to the $J/\Psi$ self-energy~\cite{psimass}.
As an example, the $DD$-loop contribution to the $J/\Psi$ self-energy is shown in Fig.~\ref{fig:loop}.

\vspace{1.5em}

To calculate the $J/\Psi$ self-energy due to the $DD, DD^*$ and $D^*D^*$ meson loops,
we use the effective Lagrangian densities
(in the following we denote by $\psi$ the field representing $J/\Psi$),
\bea
{\mathcal L}_{\psi D D} &=& i g_{\psi D D} \, \psi^\mu
\left[\overline{D} \left(\partial_\mu D\right) -
\left(\partial_\mu \overline{D}\right) D \right] ,
\label{LpsiDDbar} \\
{\cal L}_{\psi D D^*} &=& \frac{ g_{\psi D D^*}}{m_\psi} \,
\varepsilon_{\alpha\beta\mu\nu} \left(\partial^\alpha \psi^\beta\right)
\Bigl[\left(\partial^\mu \overline{D^*}^{\nu}\right) D +
\overline{D} \left(\partial^\mu D^{*\nu}\right)  \Bigr] ,
\label{LpsiDD*} \\
{\cal L}_{\psi D^* D^*} &=& i g_{\psi D^* D^*} \, \bigl\{ \psi^\mu
\left[\left(\partial_\mu \overline{D^*}^{\nu}\right) D^*_\nu
- \overline{D^*}^{\nu}\left(\partial_\mu D^*_\nu\right) \right]
\nn\\
& &\hspace{-3em}+ \left[ \left(\partial_\mu \psi^\nu\right) \overline{D^*}_\nu
- \psi^\nu \left(\partial_\mu \overline{D^*}_\nu\right)
\right] D^{*\mu}
+ \, \overline{D^*}^{\mu} \left[\psi^\nu \left(\partial_\mu D^*_\nu\right)
- \left(\partial_\mu \psi^\nu\right) D^*_\nu \right]
\bigr\} ,
\label{LpsiD*D*}
\eea
with the coupling constant value, $g_{\Psi DD} = g_{\Psi DD^*} = g_{\Psi D^*D^*} = 7.64$.
Only the scalar part contributes to the $J/\Psi$ self-energy,
and the scalar potential for the $J/\Psi$ meson
is the difference of the in-medium, $m^*_\psi$,
and vacuum, $m_\psi$, masses of $J/\Psi$,
\beq
\Delta m = m^*_\psi - m_\psi,
\label{Delta-m}
\eeq
with the masses obtained from
\beq
m^2_\psi = (m^0_\psi)^2 + \Sigma(k^2 = m^2_\psi)\, .
\label{m-psi}
\eeq
Here $m^0_\psi$ is the bare mass and $\Sigma(k^2)$ is the total $J/\Psi$
self-energy obtained from the sum of the contributions from the $DD$, $DD^*$ and
$D^*D^*$ loops. The in-medium mass, $m^*_\psi$, is obtained likewise, with the
self-energy calculated with medium-modified $D$ and $D^*$ meson masses.

In the calculation we use phenomenological
form factors to regularize the self-energy loop integrals
following a similar calculation for the $\rho$
self-energy~\cite{Leinweber:2001ac}:
\begin{equation}
u_{D,D^*}(\q^2) =  \left(\frac{\Lambda_{D,D^*}^2 + m^2_\psi}
{\Lambda_{D,D^*}^2 + 4\omega^2_{D,D^*}(\q)}\right)^2.
\label{uff}
\end{equation}
For the vertices $J/\Psi{DD},J/\Psi{DD^*}$ and $J/\Psi{D^*D^*}$,
we use the form factors 
$F_{DD}(\q^2)  = u^2_{D}(\q^2)$,
$F_{DD^*}(\q^2) = u_D(\q^2) \, u_{D^*}(\q^2)$, and
$F_{D^*D^*}(\q^2) = u^2_{D^*}(\q^2)$, respectively,  
with $\Lambda_{D}$ and $\Lambda_{D^*}$ being the cutoff masses, and
the common value, $\Lambda_{D}=\Lambda_{D^*}$ is used.

\begin{figure}[htb]
  \includegraphics[height=.35\textheight,angle=-90]{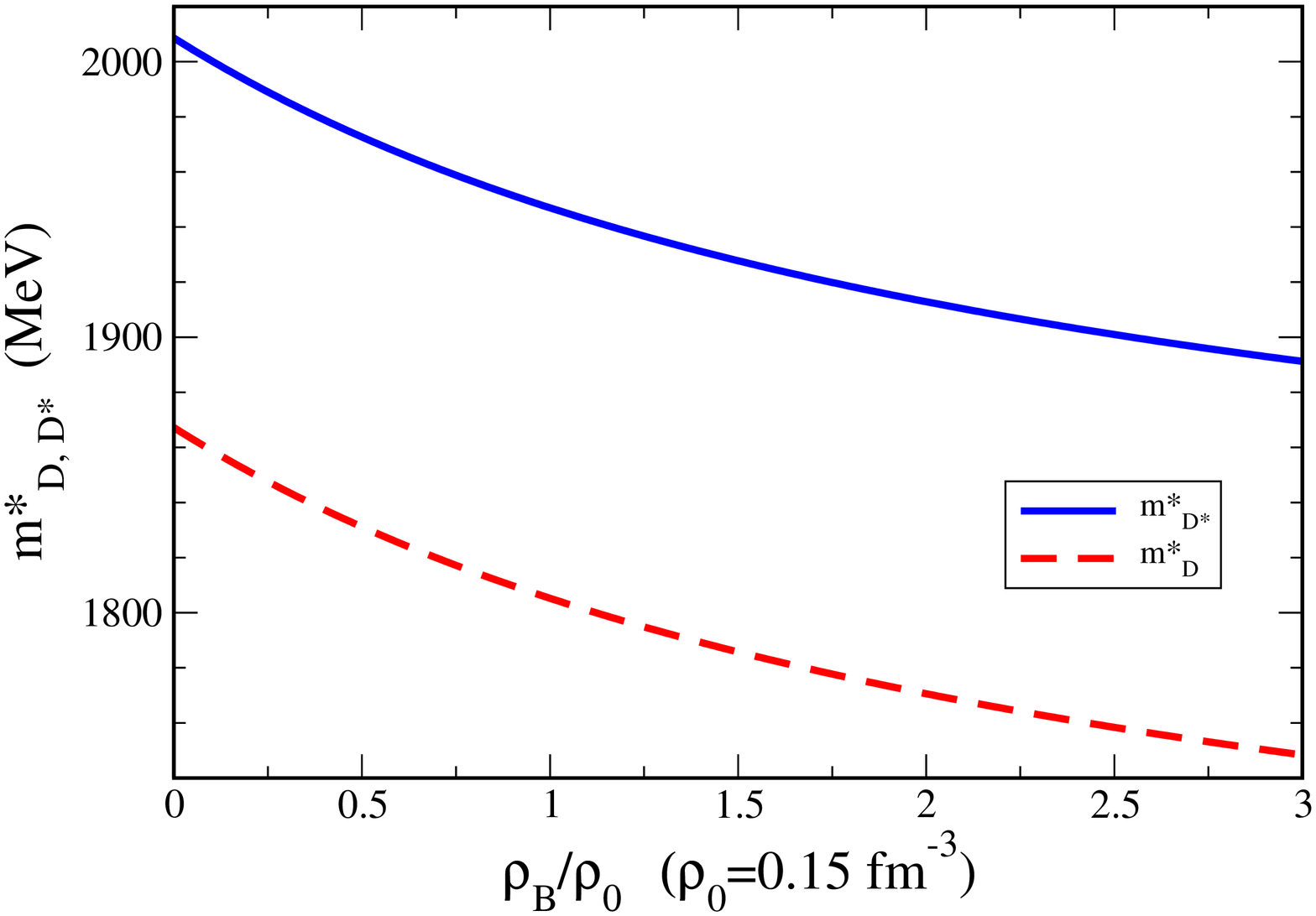}
  \includegraphics[height=.35\textheight,angle=-90]{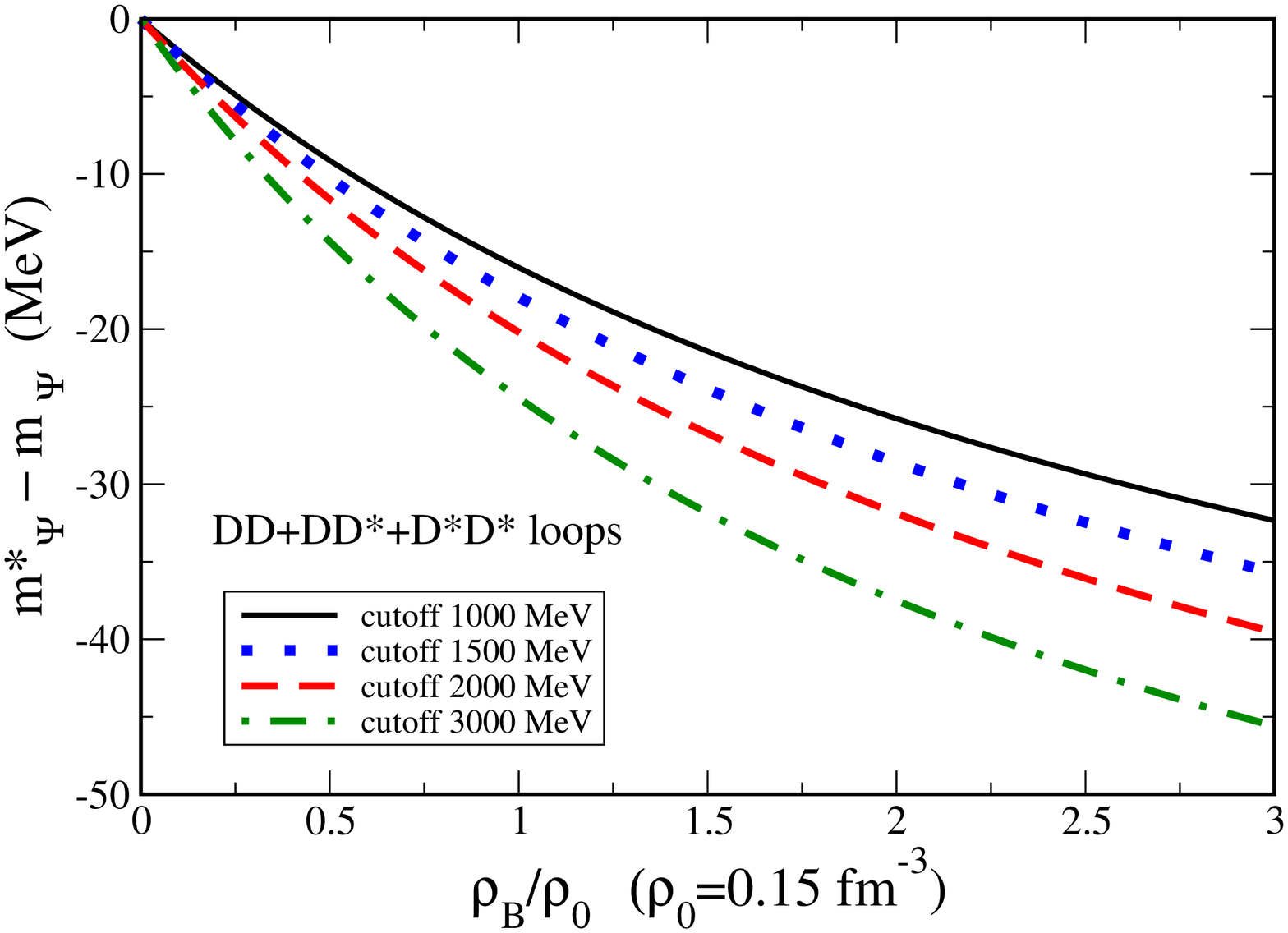}
  \caption{In-medium masses of the $D^*$ (the solid line) and $D$ (the dashed line)  
mesons calculated in the QMC model (left panel), and the $J/\Psi$ potential 
in symmetric nuclear matter (right panel). In the right panel the solid, dotted, 
dashed, and dash-dotted curves correspond to the common
cut-off values in the dipole form factors in Eq.~(\ref{uff}),
1000, 1500, 2000, and 3000 MeV, respectively.}
\label{fig:psipot}
\end{figure}

To calculate the in-medium $J/\Psi$ self-energy arising from 
the $DD, DD^*$ and $D^*D^*$ meson loops, we need to include the in-medium masses
of $D$ and $D^*$ mesons consistently~\cite{psimass,Tsushima:1998ru}.
For this purpose, we rely on the quark-meson coupling (QMC) model~\cite{Guichon,QMCreview}.
The QMC model is a quark-based, relativistic mean field model of nuclear
matter and nuclei~\cite{Guichon,QMCreview}. Relativistically moving confined light quarks in
the nucleon bags self-consistently interact directly with the scalar-isoscalar $\sigma$,
vector-isoscalar $\omega$, and vector-isovector $\rho$ mean fields generated by the
light quarks in the (other) nucleons. These meson mean fields are responsible for the nuclear binding.
The direct interaction between the light quarks and the scalar $\sigma$ field
is the key of the model, which induces the {\it scalar polarizability} at the nucleon level,
and generates the nonlinear scalar potential (effective nucleon mass), or
the density ($\sigma$-filed) dependent $\sigma$-nucleon coupling.
This gives a novel, new saturation mechanism for nuclear matter.
The model has opened tremendous opportunities for the studies of finite nuclei
and hadron properties in a nuclear medium (nuclei), based on the quark degrees of freedom.
Many successful applications of the model can be found in Ref.~\cite{QMCreview}.

\begin{table}[htb]
\begin{tabular}{cc|ccc|c}
\hline
$\;\Lambda_D\;$  & $\;m^*_{J/\Psi}\;$  & $\;DD\;$
& $\;DD^*\;$ & $\;D^*D^*\;$ & $\;\Delta m\;$ \\
\hline
                         $1000$ & $3081$ &  $-3$   & $-2$   & $-11$   & $-16$ \\
                         $1500$ & $3079$ &  $-3.5$ & $-2.5$ & $-12$   & $-18$ \\
                         $2000$ & $3077$ &  $-4$   & $-3$   & $-13$   & $-20$ \\
                         $3000$ & $3072$ &  $-6.5$ & $-5$   & $-12.5$ & $-24$ \\
\hline
\end{tabular}
\caption{In-medium $J/\Psi$ mass $m^*_{J/\Psi}$ and the individual loop
contributions to the mass difference $\Delta m$ at nuclear matter density,
for different values of the cutoff $\Lambda_D (=\Lambda_{D^*})$. All quantities are in MeV.
}
\label{tab1}
\end{table}

In Fig.~\ref{fig:psipot} we show the in-medium masses of the $D^*$ and $D$
mesons (left panel) calculated in the QMC model~\cite{psimass,Tsushima:1998ru}, 
and the $J/\Psi$ potential in symmetric
nuclear matter (right panel). To see the ambiguity due to the cut-off values
in the form factors, we calculate the potential with the cut-off values,
$\Lambda_D = \Lambda_{D^*} = 1000, 1500, 2000$, and $3000$ MeV.

In addition we list each meson loop contribution at nuclear matter density 
$\rho_0$ ($= 0.15$ fm$^{-3}$) in table~\ref{tab1}.
We regard the results with the cut-off values 1500 and 2000 MeV as our predictions.
At normal nuclear matter density, these correspond to about $18$ and $20$ MeV
attractions, respectively.

\subsection{$J/\Psi$-nuclear bound state}

\begin{figure}[htb]
  \includegraphics[height=.35\textheight,angle=-90]{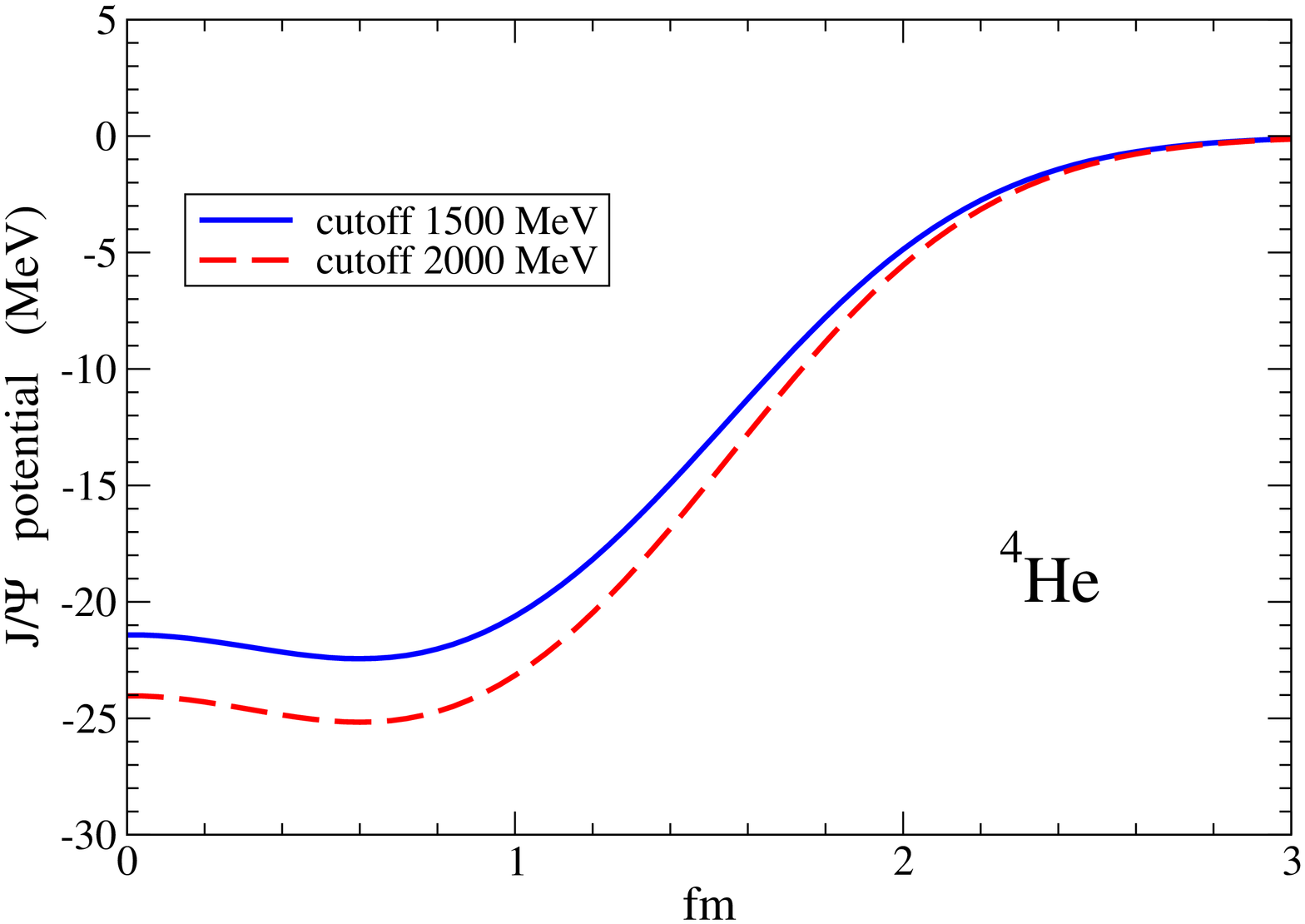}
  \includegraphics[height=.35\textheight,angle=-90]{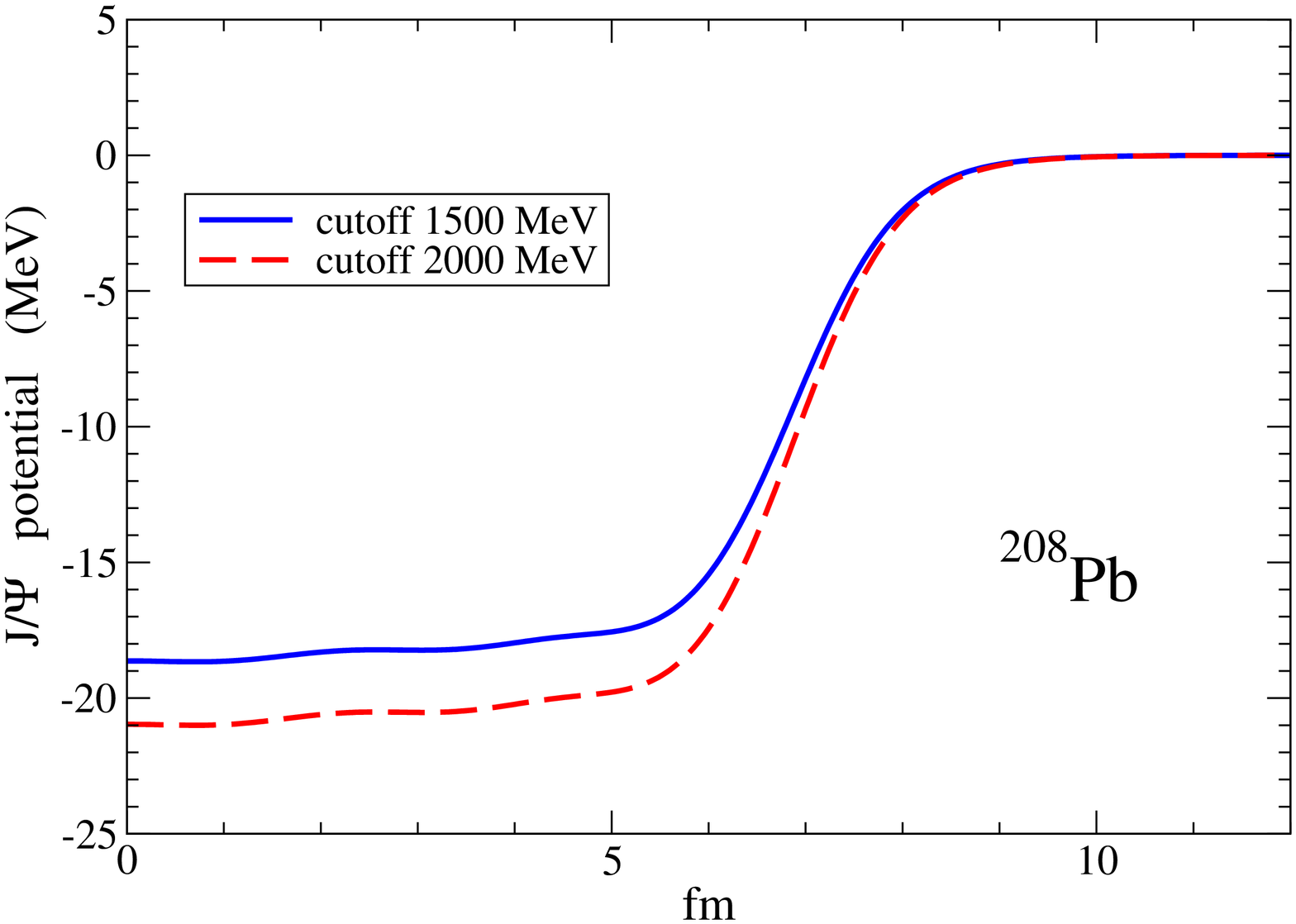}
\caption{
$J/\Psi$ potentials in $^4$He (left panel)
and $^{208}$Pb (right panel) nuclei for two
values of the cutoff values in the form factors,
$1500$ (the solid line) and $2000$ (the dashed line) MeV.
}
\label{psiApot}
\end{figure}

First, we show the $J/\psi$-nuclear potentials calculated~\cite{psimesic} based on the
method described in the previous section.
As examples, we show in Fig.~\ref{psiApot} the potentials felt by the $J/\Psi$
meson in $^4$He (left panel) and $^{208}$Pb (right panel) nuclei.

The nucleon density distributions for nuclei ($^{208}$Pb nucleus in this report) 
are also calculated within the QMC model~\cite{Saito:1996sf}.
For a $^4$He nucleus, we use the parametrization for the density
distribution obtained in Ref.~\cite{Saito:1997ae}.
Using a local density approximation
we have calculated the $J/\Psi$ potentials in nuclei.
With the potentials obtained in this manner,
we then calculate the $J/\Psi$-nuclear bound state energies.
The detail and complete results will be reported elsewhere~\cite{psimesic}.

We follow the procedure applied in the earlier work~\cite{etaomega}
for the $\eta$- and $\omega$-nuclear bound states.
In this study we consider the situation that
the $J/\Psi$ meson is produced nearly at rest (recoilless kinematics in experiments).
Then, it should be a very good approximation
to neglect the possible energy difference between the longitudinal and transverse
components~\cite{saitomega} of the $J/\Psi$ wave function.
In addition we consider the $J/\Psi$ meson is produced in medium and heavy mass nuclei,
$^{40}$Ca, $^{90}$Zr and $^{208}$Pb, where possible center-of-mass corrections
should be small. Thus, after imposing the Lorentz condition,
to solve the Proca equation, aside from a possible width,
is equivalent to solve the Klein-Gordon equation,
\bge
\left[ \nabla^2 + E^2_\Psi - m^{*2}_\Psi(r) \right]\,
\phi_\Psi(\r) = 0,
\label{kgequation}
\ene
where $E_\Psi$ is the total energy of the $J/\Psi$ meson, and $m^*_\Psi(r)$ is the
in-medium mass of the $J/\Psi$ in a nucleus.
Since in free space the width of $J/\Psi$ meson is $\sim 93$ keV~\cite{PDG},
we can ignore this tiny natural width.
Thus, we may simply solve the Klein-Gordon equation
Eq.~(\ref{kgequation}) without worrying about the width,
under the situation we consider now.
We list in table~\ref{psienergy} the bound state energies obtained.
The results show that the $J/\Psi$ meson is expected to form nuclear bound states.

\begin{table}[htbp]
\caption{
Bound state energies obtained for the $J/\Psi$ meson.
With the situation that the $J/\Psi$ meson is produced
in recoilless kinematics, the width due to the strong
interactions is all set to zero,
as well as its natural width of $\sim 93$ keV.
}
\label{psienergy}
\begin{tabular}[t]{lc|c||c}
\hline
 & &$\Lambda_{D,D^*}=1500$ MeV &$\Lambda_{D,D^*}=2000$ MeV\\
\hline
 & &$E$ (MeV) &$E$ (MeV)\\
\hline

%
%
%
%
%
%
%
$^{40}_\Psi$Ca &1s &-15.15 &-17.42\\
               &1p &-11.20 &-13.32\\
               &1d &-6.91  &-8.85 \\
%
\hline
$^{90}_\Psi$Zr &1s &-16.40 &-18.69\\
               &1p &-13.92 &-16.14\\
               &1d &-11.08 &-13.21\\
%
\hline
$^{208}_\Psi$Pb &1s &-16.80 &-19.06\\
                &1p &-15.34 &-17.57\\
                &1d &-13.62 &-15.81\\
%
\hline
\end{tabular}
\end{table}
%

\subsection{Summary and Conclusion}

We have estimated mass shift of the $J/\Psi$ meson in nuclear matter 
arising from the modification of $DD, DD^*$ and $D^*D^*$ meson loop
contributions to the $J/\Psi$ self-energy, consistently including
the in-medium masses of the $D$ and $D^*$ mesons
calculated in the quark-meson coupling model.
Then, we have calculated the $J/\Psi$-nuclear potentials using
a local density approximation, where
the nuclear density distributions are also calculated within the quark-meson
coupling model. We emphasize that, in the model, all the coupling constants between
the applied meson mean fields and
the light quarks in the nucleon, as well as those in the $D$ and $D^*$ mesons,
are all equal and calibrated by the nuclear matter saturation properties.
Using the $J/\Psi$-nuclear potentials obtained in this manner,
we have solved the Klein-Gordon equation which is reduced
from the Proca equation, and obtained the $J/\Psi$-nuclear bound state energies
for $^{40}$Ca, $^{90}$Zr and $^{208}$Pb nuclei, where possible center-of-mass
corrections and in-medium width, which are expected to be small, are all neglected.
Our results show that the $J/\Psi$ meson is expected to form nuclear bound states.

In the future we need to study the situation that
the $J/\Psi$ meson is produced with a finite momentum.
In this case the $J/\Psi$ self-energy in nuclear medium, applied
to calculate the $J/\Psi$ potential, will have an imaginary part.
Then, $J/\Psi$ meson naturally will get a momentum or energy dependent
finite width. However, these issues, momentum and energy dependence of the
effective mass or potential in nuclear medium,
as well as the momentum and energy dependence of the width,
are challenging problems which have not yet been studied well.
In particular, when one considers
the situation that the $J/\Psi$ meson propagates in the nuclear
medium (a nucleus), these issues are inevitable to be studied. 
All these are future challenging problems that we need to study.


\begin{theacknowledgments}
The work of DHL was supported by Science Foundation of Chinese University.
The work of GK was partially financed by CNPq and FAPESP (Brazilian
agencies), and AWT is supported by an Australian Laureate Fellowship
and the University of Adelaide.
\end{theacknowledgments}



\bibliographystyle{aipproc}   


%


%
\end{document}